\documentclass[pre,twocolumn,aps,superscriptaddress, nofootinbib]{revtex4-1}
\usepackage{amsmath}
\usepackage{amssymb}
\usepackage{graphicx}
\usepackage[usenames,dvipsnames]{color}
\usepackage{braket}
\usepackage{bm}
\usepackage{times}
\usepackage{hyperref}
\usepackage{pdfpages}
\usepackage{float}
\usepackage{lipsum}
\usepackage[dvipsnames]{xcolor}
\usepackage{comment}
\hypersetup{
  colorlinks=true,
  citecolor=blue,
  linkcolor=blue,
  urlcolor=blue}
\usepackage{tikz}
\usetikzlibrary{decorations.markings}
\usetikzlibrary{decorations.pathmorphing,snakes}
\usetikzlibrary{arrows.meta}
\tikzset{middlearrow/.style={
        decoration={markings,
            mark= at position 0.5 with {\arrow{#1}} ,
        },
        postaction={decorate}
    }
}

\def\be{\begin{equation}}
	\def\ee{\end{equation}}
\def\bea{\begin{eqnarray}}
	\def\eea{\end{eqnarray}}
\def\bfg{\begin{figure}[H]}
	\def\efg{\end{figure}}

\makeatletter
 \AtBeginDocument{\let\LS@rot\@undefined}
\makeatother

\begin{document}

\title{{Topological phase transition in anti-symmetric Lotka-Volterra doublet chain}}

\author{Rukmani Bai$^+$}
\email{rukmani.bai@itp.uni-hannover.de}
\affiliation{Institut f\"ur Theoretische Physik, Leibniz Universit\"at Hannover,
Appelstrasse 2, D-30167 Hannover, Germany}
\affiliation{Institute for Theoretical Physics III and Center for 
             Integrated Quantum Science and Technology, University of 
	     Stuttgart, 70550 Stuttgart, Germany}

\author{Sourin Chatterjee$^+$}
\affiliation{Department of Mathematics Statistics, Indian Institute of Science Education and 
             Research, Kolkata, West Bengal 741246, India}
\affiliation{Institut de Neurosciences des Systèmes (INS), UMR1106, Aix-Marseille Université, Marseilles, France}             

\author{Ujjwal Shekhar}
\affiliation{Center for Computational Natural Sciences and Bioinformatics, International Institute of Information Technology, Gachibowli, Hyderabad 500032, India}
             
\author{Abhishek Deshpande}
\affiliation{Center for Computational Natural Sciences and Bioinformatics, International Institute of Information Technology, Gachibowli, Hyderabad 500032, India}

\author{Sirshendu Bhattacharyya}
\email{sirs.bh@gmail.com}
\affiliation{Department of Physics, Raja Rammohun Roy Mahavidyalaya, Radhanagar, Hooghly 712406, India}
\author{Chittaranjan Hens}
\affiliation{Center for Computational Natural Sciences and Bioinformatics, International Institute of Information Technology, Gachibowli, Hyderabad 500032, India}

\def\thefootnote{+}\footnotetext{These authors contributed equally to this work.}

\begin{abstract}
We present the emergence of topological phase transition in the minimal model of two dimensional rock-paper-scissors cycle in the form of a doublet chain. The evolutionary dynamics of the doublet chain is obtained by solving the anti-symmetric Lotka-Volterra equation. We show that the mass decays exponentially towards edges and robust against small perturbation in the rate of change of mass transfer, a signature of a topological phase. For one of the configuration of our doublet chain, the mass is transferred towards both edges and the bulk is gaped. Further, we confirm this phase transition within the framework of topological band theory. For this we calculate the winding number which change from zero to one for trivial and a non-trivial topological phases respectively.  
\end{abstract}

\maketitle
{\it Introduction.}
Topology plays a vital role  in condensed matter physics \cite{thouless1982,jiang2012fermi,haldane2015}.
Topological phase transition is one of those properties that has been previously studied in the context of quantum systems, but recent works have shown this phenomenon is observed in mechanical meta-materials\cite{kane2014topological,chen2016,fan2019}, active matter \cite{sone2020exceptional}, and photonic crystals \cite{raghu2008,ozawa2019,he2020quadrupole}. Robustness, condensation, localization and phase transitions are key characteristics of this kind  of system~\cite{haldane2015}. This condensation phenomenon has also been observed in systems like citation networks~\cite{krap2000,bianconi2001}, jamming of traffic \cite{kappu2005} or mass transport models\cite{evans2014condensation}. As biological systems and soft matter have shown properties of topological phases \cite{prodan2009, nash2015topological, souslov2017topological, pedro2019,yoshida2021bulk}, one  key area of interest is how these non-trivial topological properties can be obtained in a designed biological set-up.

Recent studies have shown that anti-symmetric Lotka-Volterra Equation (ALVE) applied to a lattice or chain-like structure can essentially reproduce the properties of topological phase \cite{knebel2020topological,yoshida_22,liang_24} or chiral edge modes\cite{yoshida2021chiral}. ALVE is a non-linear mass-preserving model of entities interacting in cycles, which has been introduced for the study of population dynamics. In a generic system of $S$ constituents, mass can be regarded as the normalized quantity of each of them. Even if the constituents are interacting with each other and their mass is being transferred, the total mass remains constant throughout the process. Interestingly, ALVE has been used in different fields like quantum physics \cite{vorberg2013generalized}, evolutionary game theory \cite{chawanya2002large}, population dynamics \cite{goel1971vol, may2019stability}, chemical kinetics \cite{di1989limit} and plasma physics \cite{zakharov1974nonlinear, manakov1975complete} to describe condensation processes. This provides a motivation to establish equivalence between such systems. Recently, Knebel {\it et.al.}\cite{knebel2020topological} have shown how the properties of an ALVE equation applied on a one-dimensional Rock-Paper-Scissor (RPS) chain on the basis of localization and robustness,  can be used to mimic the topological phase transition. Particularly, with a suitable choice of interaction, mass is transferred largely to the boundary of the chain. The result is  robust with respect to the model parameter heterogeneity, and a phase transition occurs at a critical value of the parameter.

Mapping RPS systems beyond one dimensional lattices have also been exercised earlier. For example, an RPS system within the structure of Kagome lattice have shown the emergence of the chiral edge mode \cite{yoshida2021chiral}. RPS interaction realized in 3D lattice has been reported to exhibit surface polarization of mass that can be understood from the 3D Weyl semimetal phases \cite{umer_22}. These works have opened up avenues to understand dynamical features of one and higher dimensional nonlinear systems in the light of topological band theory which was earlier limited to linear systems only.

\begin{figure}[t]
    \centering
   \includegraphics[width=0.5\textwidth]{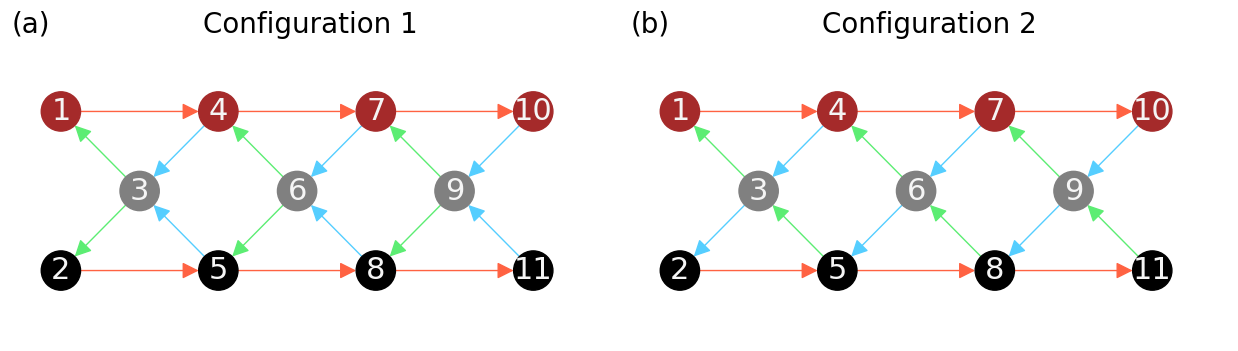}
    \caption{ Doublet chain of Rock-Paper-Scissor cycle. The arrows from one node to another indicate the mass transferring with a specific rate $r_1$, $r_2$, and $r_3$.
    red: $r_1$, blue: $r_2$, and green: $r_3$. These four configurations are named (a) Configuration 1, (b) Configuration 2.}
    \label{fig-network}
\end{figure}

With similar motivation i.e. to investigate the topological aspect of the RPS system in higher dimensions, we start with a doublet chain as an extension and explore the mass transfer propagation from one node to other through ALVE dynamics. This doublet chain is constructed by periodic repetition of two merged RPS triangles via a single vertex  (See Fig.\ \ref{fig-network}). Keeping the interaction of the RPS chain similar to the one in~\cite{knebel2020topological}, we found that the masses are shifted to one side of the doublet chain; a property similar to what was observed in~\cite{knebel2020topological}. Moreover, as this structure can encompass more complexity by keeping one RPS chain as before and swapping the interaction between Paper-Scissor and Scissor-Rock, most of the masses are concentrated at the diagonal corner which has not been observed before. These results are robust in terms of parameter values and remain present for large doublet chains.

{\it Model.}
The ALVE describing the mass at each site $\alpha$ is given by a system of coupled non-linear ordinary differential equations, 
\bea 
\frac{dx_{\alpha}}{dt}=x_{\alpha}\sum_{\beta = 1}^{S}a_{\alpha \beta}x_{\beta}.
\eea
where $\alpha=1,2,\cdots S$, and $a_{\alpha \beta}$ are the elements of the $S \times S $ antisymmetric matrix $A$. 
The site $\alpha$ and its interaction with its neighbours are determined by the elements of $A$. Particularly, $a_{\alpha \beta}$ is considered as the (signed) rate at which mass is transferred from the site $\beta$ to site $\alpha$.
Here the whole system conserves mass at each time ($ \sum_{\alpha = 1}^{S} x_{\alpha}(t) = \text{Const.} \; \forall \: t $) \cite{knebel2020topological}. To start with, we have taken normalized ($\sum_{\alpha = 1}^{S} x_{\alpha}(t=0) = 1 $), random, and non-zero initial masses ($x_{\alpha}(t=0) \geq 0 \; \forall \: \alpha $) in the system.

\par 
To take a step towards understanding the two dimensional system, we extended the 1-D chain proposed by Knebel {\it et al} \cite{knebel2020topological} to a doublet structure as shown in Fig.~\ref{fig-network}. Here, we construct it by joining two triangle sub-units through one of the vertex and then extending it by repeating. On each triangle sub-unit, mass transfer is cyclic. For both configurations, red, blue, and green, arrows represent the strength of  $r_1$, $r_2$, and $r_3$ respectively. The direction of the arrow represents the direction of mass transfer. Here we have shown two types of doublet configurations depicted in Fig.\ \ref{fig-network}.
We create these two configurations by switching between $r_2$ and $r_3$ in the lower triangular chain.     
For configurations 1 \& 2, the flow of mass in the upper nodes and lower nodes are in the same direction. Hence, following the direction of the arrows, we find anti-clockwise motion in the lower triangles and clockwise motion in the upper triangle. For configuration 1, the middle nodes (numbered as $3,6,9,..$) will have  4 connections, in which these nodes have two incoming links of strength $r_2$ (blue) and two outgoing links of strength $r_3$ (green). However, for configuration 2, these middle nodes have one incoming and outgoing link having the strength of $r_2$ (blue) and $r_3$ (green) each. No periodic boundary conditions have been applied here.
The corresponding matrix for this configuration 1 is as follows:

\bea
A_{1}=
\begin{bmatrix}
0 & 0 & r_3 & -r_1 & 0 & 0 & 0 & 0 & \dots & 0 \\
0 & 0 & r_3 & 0 & -r_1 & 0 & 0 & 0 & \dots & 0\\
-r_3 & -r_3 & 0 & r_2 & r_2 & 0 &  0 & 0 & \dots & 0\\
r_1 & 0 & - r_2 & 0 & 0 &  r_3 & -r_1 & 0   & \dots & 0 \\
0 & r_1& -r_2 & 0 & 0 & r_3 & 0 & -r_1 &  \dots & 0\\
0 & 0 & 0 & -r_3 & -r_3 & 0 & r_2 & r_2 & \dots & 0 \\
\vdots & \vdots & \vdots   & \vdots & \vdots & \ddots & \vdots & \vdots   & \vdots & \vdots  \\
0 & 0 & 0 & 0 & 0 & 0 & \dots & 0 & -r_3 & -r_3\\
0 & 0 & 0 & 0 & 0 & 0 & \dots & r_3 & 0 & 0\\
0 & 0 & 0 & 0 & 0 & 0 & \dots & r_3 & 0 & 0\\
\end{bmatrix}.
\nonumber
\eea
The matrix $A_2$ corresponding to configuration 2 is given in the Appendix \ref{app1}. The ratio $\frac{r_2}{r_3}$ can be referred to as a skewness parameter ($r$) as control over it determines the mass condensation phenomenon to a particular orientation. 
 Each of the simulations is performed by generating 600,000-time points data using an adaptive ODE45 solver. The last 160,000-time points data have been used to calculate the average mass density ($\langle x_{\alpha} \rangle_T = \frac{1}{T} \int_{0}^T x_{\alpha}(t) dt $). 
 The simulations are also cross-checked with Runge-Kutta 4 routine. We have observed similar results. 

\vspace{2pt}
Here, we analyze the data for a graph of $S=23$ nodes to state the result with visual representation. The results for the larger graph are shown in the appendix \ref{app3}. 
We have divided the structure into three chains- the upper chain (i.e. node index 1, 4, 7, etc.) is marked with brown color. The middle chain (i.e. node index 3,6,9...), and the lower chain (i.e. node index 2,5,8...) are marked with gray and black 
color respectively. 
Throughout our investigation, this colour as well as the labelling scheme were maintained. 

{\it Configuration 1.}
\label{Configuration 1}.
\begin{figure*}
	\centering
    \includegraphics[scale=0.37]{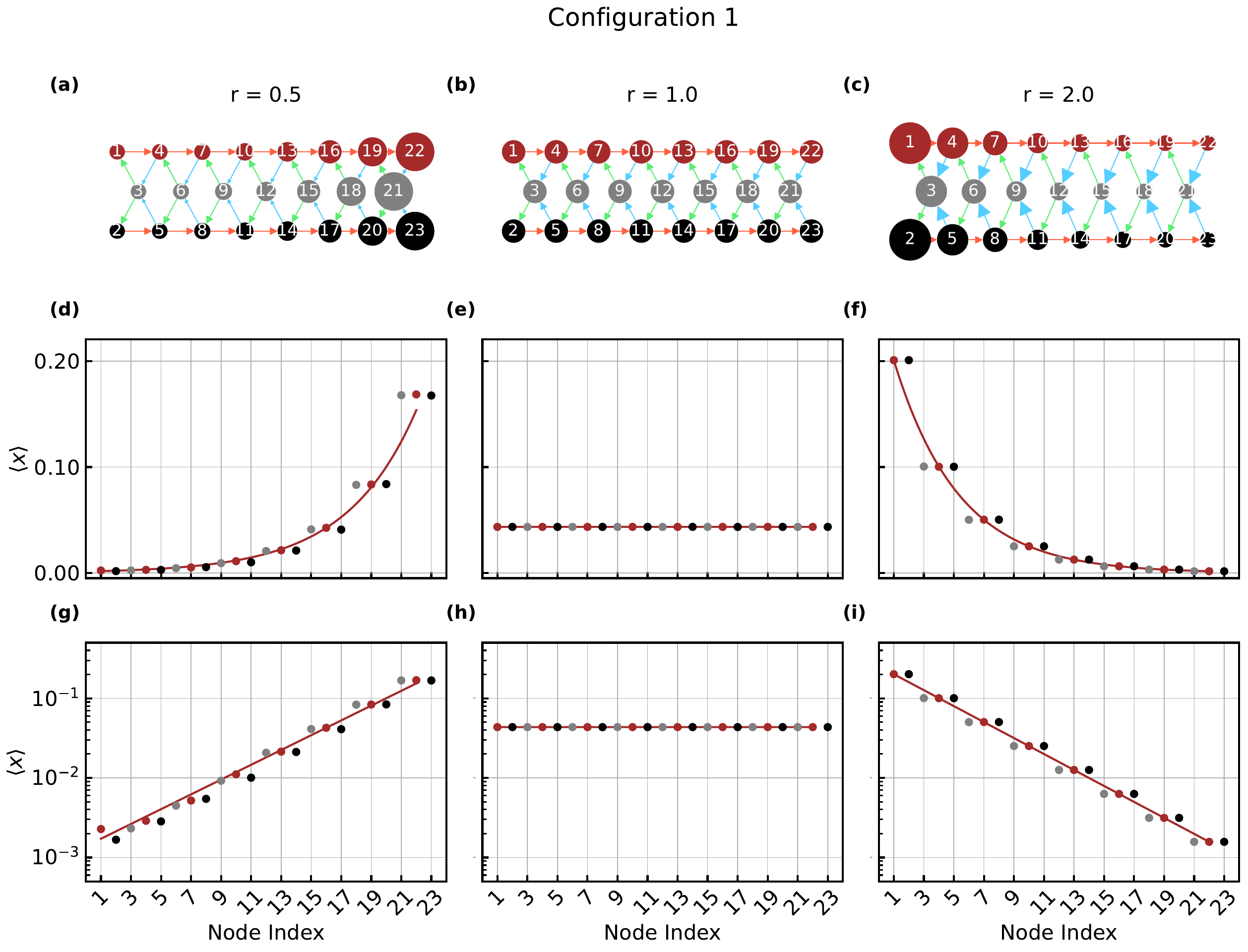}
	\caption{Node density averaged over time, corresponding to Configuration 1. A visual representation of the network is given here. The strength of the rate constant is represented by the width of the arrow. The node density is represented by the size of the node. (d-f) The average density of nodes are plotted in linear scale where as for more clarity we have used semi-log scale in (g-i). The exponential decline of mass deposition towards one end is clearly visible. 
    (a) and (d) corresponds to $r_2 = 0.5$, (b) and (e) corresponds to $r_2 = 1$, (c) and (f) corresponds to $r_2 = 2$.  The straight lines (left and right)   reconfirm  the exponential nature of the mass distribution. }. 
	\label{fig-2}
\end{figure*}
At first, we consider the case for 
$r= 0.5$ ($r_1=1;  ~ r_2=0.5; ~ r_3=1$). Here, the strength of each blue link is weak compared to the others (red or green links). We observe that most of the masses are transferred to the right side of the doublet chain i.e.at node index 21, 22, and 23 (see Fig.\  \ref{fig-2}(a), (d)). The density of the mass is highest in the right and then drops towards the left. This decays exponentially (Fig.\ \ref{fig-2}(g), where the simulated data is replotted in semi-log scale, makes this more evident).   
This result is analogous to earlier works in 1-d chain \cite{knebel2020topological} where mass is polarized in the right boundary when $r<1$. Considering the grey nodes in this scenario, the incoming blue links are weaker than the green links. Mass can thus be transferred with ease to the upper  (brown nodes) or lower surface (black nodes). Since red and green links are equally strong, all mass is progressively moved in the direction of the red links and ultimately deposited on the right side.
Now, we take all the parameters as identical, i.e. $r=1$ ($r_1=r_2=r_3=1$). 
Here, we find that the masses get uniformly distributed throughout the network (see Fig.\ \ref{fig-2}(b), (e), and (h) (semi log-scale)), as all rate constants are equal.
Finally, in the case of 
$r = 2.0$ ($r_1=1.0;~ r_2=2.0; ~ r_3=1.0$), 
we observe that the average masses are highly concentrated on the left side (opposite direction to that of when $r=0.5$) i.e.node indexes 1,2 and 3 have a higher average density compared to others  (see Fig.\ \ref{fig-2}(c),(f), and (i)). The explanation is as follows: because the blue links are stronger than the green and the red links, the mass eventually shifts to the left under the influence of the direction of blue links. 
Here the masses are polarized in one direction, a consistent pattern with the previous work \cite{knebel2020topological}. It is also clear, that the highest-density nodes are $\sim 100$ times bigger than the lowest-density nodes ($r = 0.5$ as well as $r = 2$). Thus, the decay constant is almost the same, as the value of $r$ in one case is reciprocal to the other case.


\begin{figure*}
	\centering
	\includegraphics[scale=0.37]{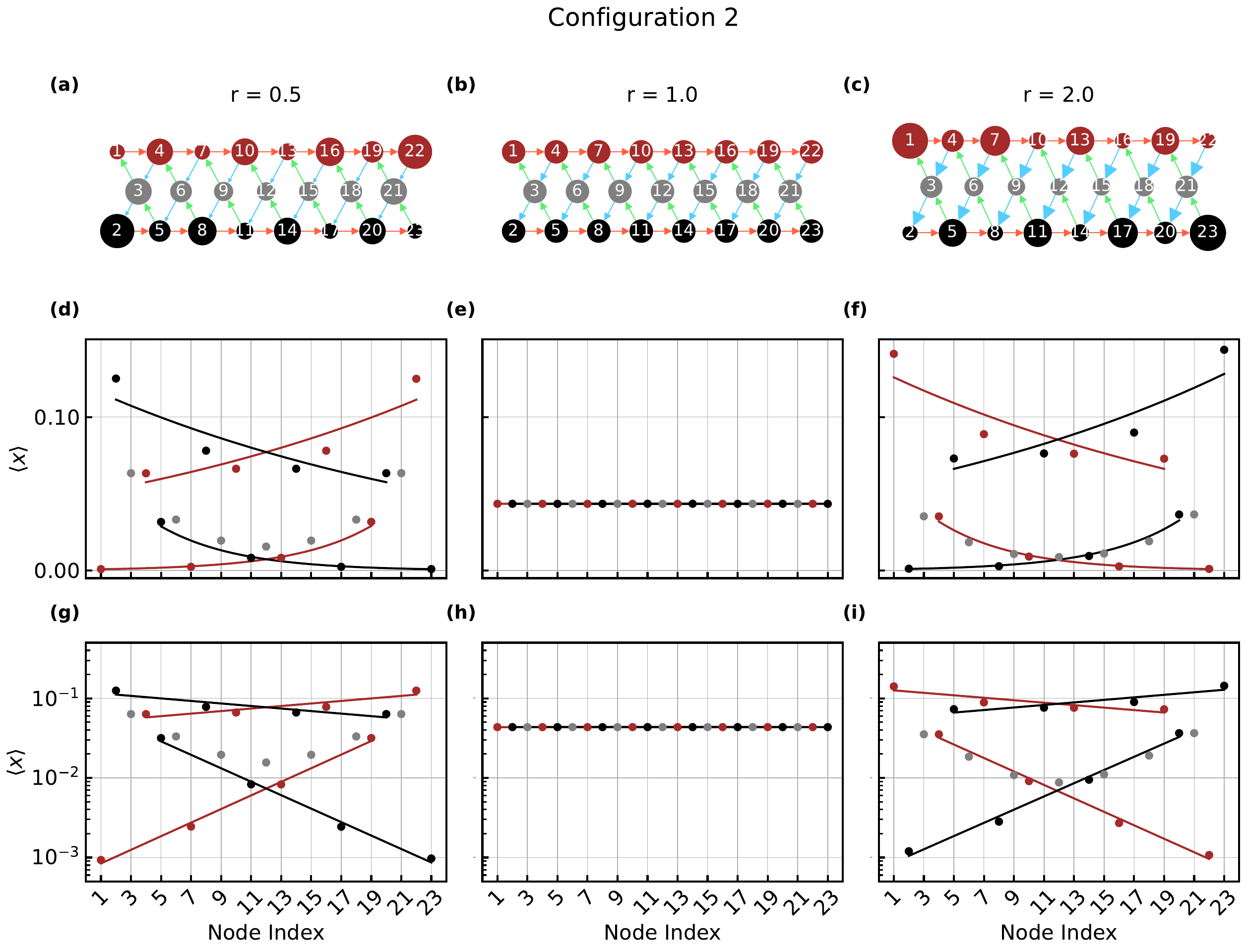}
	\caption{Node density averaged over time, corresponding to Configuration 2.  The strength of the rate constant is represented by the width of the arrow. The node density is represented by the size of the node. (d-f) Each node's  average density is reported in linear scale. (g-i) Node density is shown in semi-log scale. There are two decays rates of brown/black nodes. Depending on the rate strength ($r>1, {~\rm or ~} r<1$), the mass is deposited in opposite two corners of this doublet configurations.  
    (a), (d), and (g) correspond to $r_2 = 0.5$, (b), (e), and (h) correspond to $r_2 = 1$, (c), (f), and (i) correspond to $r_2 = 2$. }
	\label{fig-3}
\end{figure*}

\begin{figure}
	\centering
	\includegraphics[width= 0.3\textwidth]{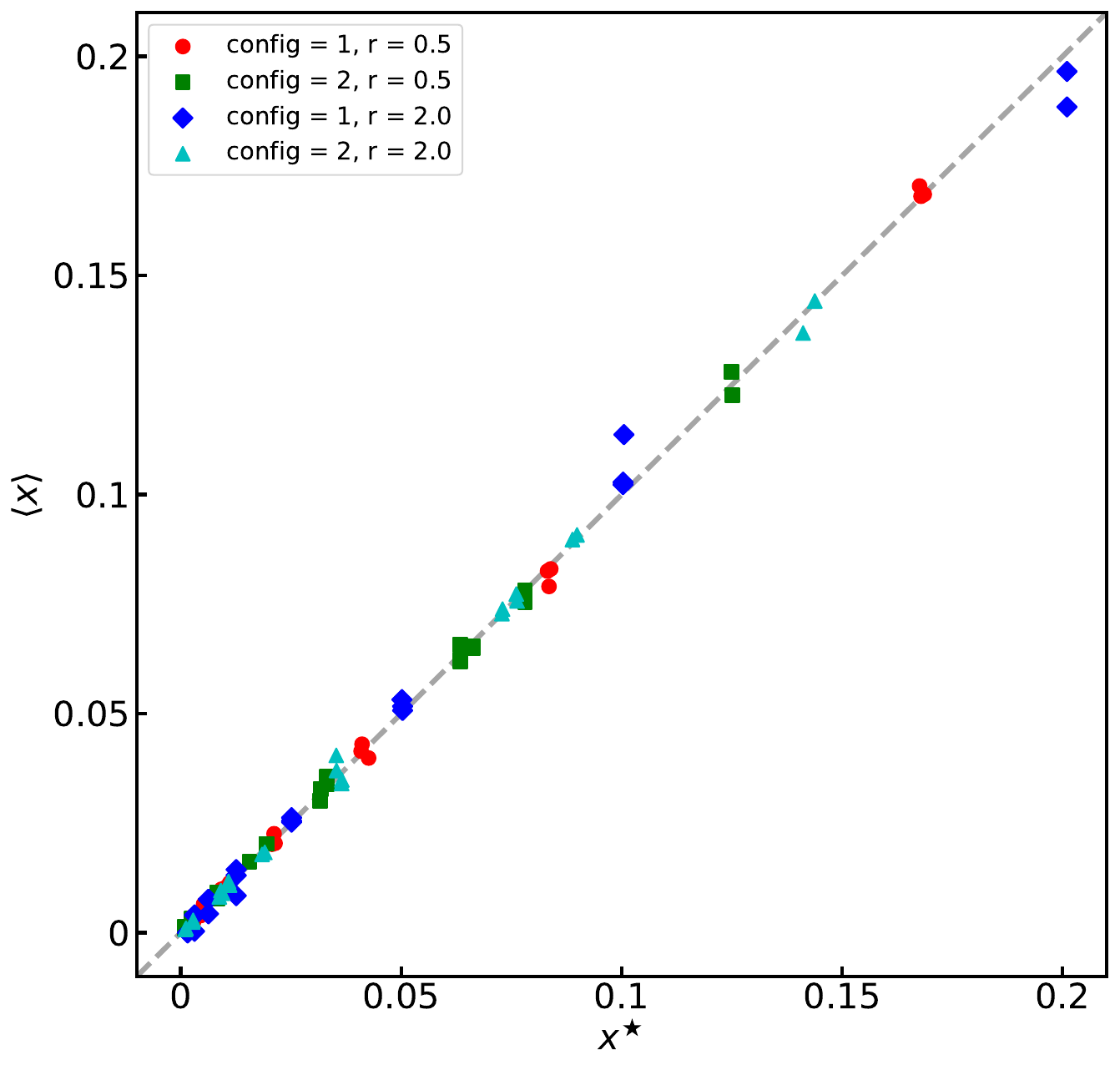}
	\caption{{We have accumulated four  average densities of each node (two parameters for each configuration), and  plotted with the value obtained from the equation $Ax=0$. Note that $A$  will also change according to the configuration and rate parameters. }}.
	 \label{fig-avgxvsx*} 
\end{figure}

{\it Configuration 2.}  Initially, $r=0.5$ ($r_1=1.0; r_2=0.5; r_3=1.0$) was taken into consideration. In comparison to configuration 1, we observe that the highest density is found in the network's two opposing corners (node index 2 and 22; see Fig.\ \ref{fig-3}(a), (d)). The mass is therefore not deposited on one side alone in that scenario. Additionally, red nodes are seen to follow two exponential decays. The mass decays exponentially through alternating nodes ($22-16-10-4$), with node 22 having the largest accumulation of mass. Exponential decay is also followed by the other alternative set of red nodes ($19-13-7-1$). Readers can examine the two red lines and the red nodes for more clarity (Fig.\  \ref{fig-3}(g)). Similar behavior, but in the opposite direction, is seen in the black nodes (mass decays exponentially from 2 to 8 to 20 via 14).
The identical rates  ($r_1=r_2=r_3=1.0$) show that the masses become uniformly distributed throughout the network from a random mass distribution (see Fig.\ \ref{fig-3}(b), (e), and (h)). This instance is identical to configuration 1's. Now, where $r=2$($ r_1=1; r_2=2; r_3=1$), as opposed to the case when $r=0.5$, we observe that masses are becoming polarized and concentrating heavily on two opposing corners of the network (node index 1 and 23) (see Fig.\ \ref{fig-3}(c), (f), and (i)). Note that  in configuration2, if  $r\neq1$, the middle of the  gray nodes deposits comparatively less mass. Contrast to configuration 1, the directionality and strength of the links in this configuration cannot explain the average mass distribution. For instance, node 10 and 13 have same link structure, although, node 10 has strong mass deposition.

To understand this behaviour more accurately, we have checked the fixed points of these configurations. The system has one trivial fixed point vector (${\mathbf{x=0}}$). Solving $A{\bf x}=0$ will yield the value of the other fixed point vector (solution: ${\bf x^*}$). Upon numerically solving it, we have plotted ($\langle x \rangle_{\alpha}$) the average density of each site with ${x_\alpha}^*$. It is shown in Fig. \ref{fig-avgxvsx*}. It can be inferred from the straight line that each node's trajectory circles the fixed point (see \cite{knebel2020topological,reichenbach2006coexistence,reichenbach2007noise,bhattacharyya2020mortality,chatterjee2022controlling,viswanathan2024ecological,chatterjee2024combined}). Additionally, we looked for a large graph ($N=53$); refer to appendix \ref{app3}. Note that, for a large graph, we consider the initial condition as a small perturbation of the solution of $A{\bf x}=0$. Particularly, we  initially perturb with a small strength to one of the nodes. As a result, relatively large oscillation emerges, however, the average behaviour of the system can be approximated with an intrinsic non-trivial fixed point vector. Noticeably, if the perturbation is too strong or initial states are chosen randomly, the average behaviour of the nodes may not be mapped with ${\bf x^*}$ (appendix \ref{app3}). 
It is evident now that when $r$ increases from less than 1 to greater than 1, the exponents will change sign. Is it possible to reveal phase transitions as a function of $r$ in both configurations? is a crucial question to pose. We have thoroughly examined it with analytical computation in the sections that follow.
Prior to the analytical approach section, we would like to point out that, in the event that the network size is even, a unique solution for $A{\bf x}=0$ does not exist. Therefore, it is possible that the pattern of odd sizes we have observed will not show up for even sizes. This subject will be covered in greater detail in the future works. As a result, only the system size $S=6n-1; ~~ n=1,2,3,...$ is covered by our investigation.

\begin{figure}
    \centering
    \includegraphics[width=0.5\textwidth]{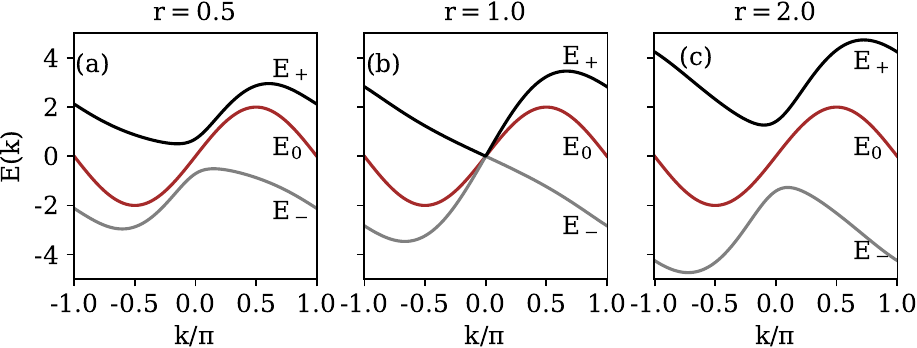}
    \includegraphics[width=0.5\textwidth]{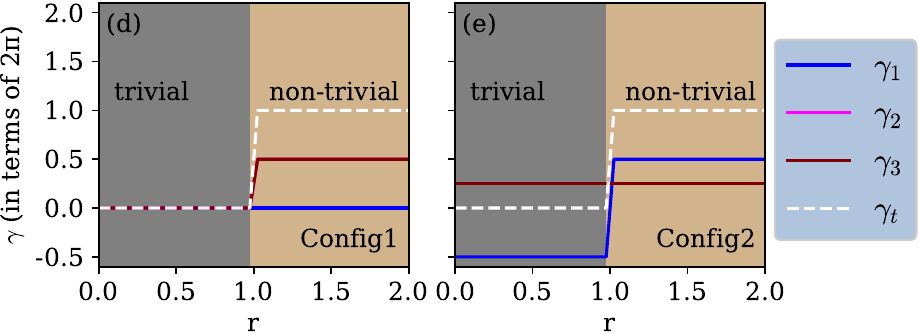}
    \caption{(a)-(c) Energy spectrum for configuration 1 and 2 as a function of rate $r$. We observe that 
    the topological phase transition occurs at $r = 1$ where energy gap between three bands closes at $k = 0$. For $r<1$ and $r>1$ the three energy bands are gaped identifying two distinct topological phases. 
    (d)-(e) The topological phase transition occurs at $r =1$ for configuration 1 and 2. The winding number is zero for the topological trivial phase ($r<1$) and it is integer one for the topological non-trivial phase ($r>1$). }
    \label{fig:ener}  
\end{figure}

{\it Analytical Approach.}
\label{Analytical Approach}
We now make use of the topological band theory to understand the topological properties of this RPS doublet chain \cite{knebel2020topological}. For this, we employ periodic 
boundary condition on the RPS doublet chain in Fig.\ \ref{fig-network}, and write the matrix $A$ as 
$A_{PBC} = Circ(A_0, A_1, 0,...,A_{-1})$ see appendix (\ref{app1}) for details. Here 
$A_0, A_1 \& A_{-1}$ are three block matrices, and for configuration 1, we have  
\begin{equation} 
A_0^{c1} = \begin{pmatrix}
0 & 0 & r_3\\
0 & 0 & r_3\\
-r_3 & -r_3 & 0
\end{pmatrix}, \,\,\,
A_1^{c1} = \begin{pmatrix}
-r_1 & 0 & 0\\
0 & -r_1 & 0\\
r_2 & r_2 & 0
\end{pmatrix},
\label{mat_acl1}
\end{equation}
with $A_1^{c1} = (-A_{-1}^{c1})^T$. Thus, $A_{PBC}$ becomes the block circulant, 
a translationally invariant matrix \cite{gray_06}. Now, to understand 
our system within the framework of topological band theory, we define the RPS doublet 
chain Hamiltonian $H := i A_{PBC}$. The Hamiltonian $H$ constructed in this way is Hermitian 
with $i$ being the imaginary part of a complex number. We now analyze the spectral properties
of this Hamiltonian $H$ and show that the topological properties of the doublet chain depend 
on the parameter $r$. Since $H$ has transnational invariance, we can do {\it plane wave decomposition} of the eigenvectors $\Tilde{u}(k)$ of $H$,
and the eigenvalue equation of $H$ has the form $\Tilde{H}(k) \Tilde{u}(k) = \lambda(k) \Tilde{u}(k)$. Here $\Tilde{H}(k)$ is the Fourier transformed Hamiltonian that will depend 
on the block matrices stated in the Eq. (\ref{mat_acl1}) and is given by 
$\Tilde{H} (k) = e^{-ik}(-H_1^T) + H_0 + e^{ik}H_1$, with $H_0 := iA_0, H_1 := iA_1, H_{-1} := iA_{-1}$ \cite{knebel2020topological} , $k$ being the wave number. 
With this, $\Tilde{H}(k)$ for configuration 1 has the form,
\begin{equation}
\Tilde{H}_{c1}(k) =
\begin{pmatrix}
 2r_1\sin(k) & 0 & ir_3 - ir_2e^{-ik}\\
    0 &  2r_1\sin(k) &  i r_3 - i r_2 e^{-i k}\\
    -ir_3 + ir_2e^{ik} &  -i r_3 + i r_2 e^{i k} & 0\\
\end{pmatrix}
\nonumber
\end{equation}
The Hamiltonian matrix for the configurations 2 is given in the appendix \ref{app1}. 
The configuration 1 and 2 have the same eigenvalues 
\begin{eqnarray}
    \begin{split}
        E_0 &= 2 r_1 \sin(k) \nonumber \\ 
        E_{\pm} &=  r_1 \sin(k) \pm \sqrt{ r^2_1 \sin^2(k) + 2(r^2_2 + r^3_3 - 2r_2 r_3 \cos(k)) }.
    \end{split}
\end{eqnarray}
Here eigenvalues $E_{\pm}$ are point symmetric with respect to origin that is $E_+(k) = -E_-(-k)$. This is due to the particle hole symmetry of the Hamiltonian $\Tilde{H}(k)$.
The energy spectrum of $\Tilde{H}(k)$ depends on the rate $r$, and shown in the Fig \ref{fig:ener}.
Since we have defined $r = r_2/r_3$, we fix rates $r_1 = r_3 = 1$ and vary the rate $r_2$. 
Then we calculate the energy spectrum $E(k)$ for different values of $r$ and plot them in 
Fig \ref{fig:ener} (a)-(c). The eigenvalue spectra of $\Tilde{H}(k)$ have three energy bands
for different $r$ on the Brillouin zone $k \in [-\pi,\pi]$. We observe that the energy gap closes at $k = 0$
as we approach $r=1$ and then reopens as we go away from that point. This is exactly 
what happens in the case of a topological phase 
transition \cite{jangjan2021topological}. We can therefore label the two phases 
separated by $r=1$ to be topologically distinct.\cite{knebel2020topological}. In 
Fig \ref{fig:ener}, variation of eigenvalues $E(k)$ is plotted against $k$ for three 
cases $r=0.5$(a), $r=1$(b), $r=2$(c). We observe that $r=1$ serves 
as a critical value where the eigenvalues meet at one point in the Brillouin zone. The other two 
cases $r=0.5$ and $r=2$, have a gap between two non-trivial eigenvalues that can be 
termed band gaps. For both configuration, we observe from the energy spectrum that we can not go from 
the case (a) to case (c) without going via case (b), which suggests case (a) and case (c) are 
two topologically distinct phases. 

Further to identify these two distinct topological phases, we calculate the topological 
invariant such as winding number. For this, we calculate the Zak phase
for each energy band \cite{zak_89}, and the winding number is given by the sum of the Zak phase for all bands
\cite{yan_23, anastasiadia_22, jiang_18, midya_18} at a particular rate $r$. The Zak phase is given by
\begin{equation}
\gamma_j = \frac{i}{2\pi}\int_{\text{BZ}} \bra{\Tilde{u}_j(k)} \partial_k \ket{\Tilde{u}_j(k)} dk
\end{equation}
where $\ket{\Tilde{u}_j(k)}$ is the eigen vector for each band ($j =1,2,3$)
\cite{xiao_10, zhang_21}. 
We calculate the winding number which is the sum of the total Zak phase $\gamma_t = \gamma_1 + \gamma_2 + \gamma_3$, shown with a white dashed line in the Fig \ref{fig:ener} (d) - (e). We observe the topological phase transition at $r=1$ where the winding number change from zero to one for both configuration. Here winding number zero indicates the topological trivial phase while winding number one represents the topological non-trivial phase.        

{\it Discussion.} We have studied topological phase transition in the doublet chain of RPS cycle. The considered model mimic the minimal set up of a 2D system. The mass transfer within this doublet chain is obtained by solving the ALVE equation. We have explored that for one of our configuration, the average masses are accumulated towards edges and decay exponentially consistent with results of one dimensional RPS chain. However for another configuration 2 the masses are deposited in the opposite corner and decays exponentially in an alternate way. With this we have observed the edges on both side of the RPS doublet chain, which is expected in a 2D system. The observations are robust for different network sizes and a wide range of rate parameters.

We further confirm this topological phase transition  by constructing a Hamiltonian within topological band theory. We show that the energy gap between the bands closes for parameter $r=1$, and for $r<1$ and $r>1$ the bands remain open. It suggests that there is a topological phase transition when system passes at $r=1$. We confirm this phase transition by calculating the winding number which is the sum of Zak phases for each band. Our observation shows that the winding number is zero for trivial topological phase ($r<1$), and one for non trivial topological phase ($r>1$).   

In  the future, it will be interesting to study the 2D system by adding more layers of RPS chain in our suggested doublet chain model. However it is a highly complex problem as there will be several configurations to construct by changing the cycle of RPS chains.   

{\it Acknowledgement.} We gratefully acknowledge fruitful discussions with Sahil Islam and Mauro Mobilia. 
R. B. acknowledges the support of the Deutsche Forschungsgemeinschaft (DFG, German Research Foundation) under
Germany’s Excellence Strategy – EXC-2123 Quantum- Frontiers – 390837967.

{\it Authors Contributions.} S.C. and U.S. performed the non-linear dynamical calculations. R.B. and S.B. performed analytical and theoretical calculations. Everyone contributed to writing the paper.

\bibliography{tpp_main}{}

\appendix
\section{Antisymmetric Matrix }
\label{app1}
\begin{figure*}[htbp]
	\centering
	\includegraphics[width= 0.8\textwidth]{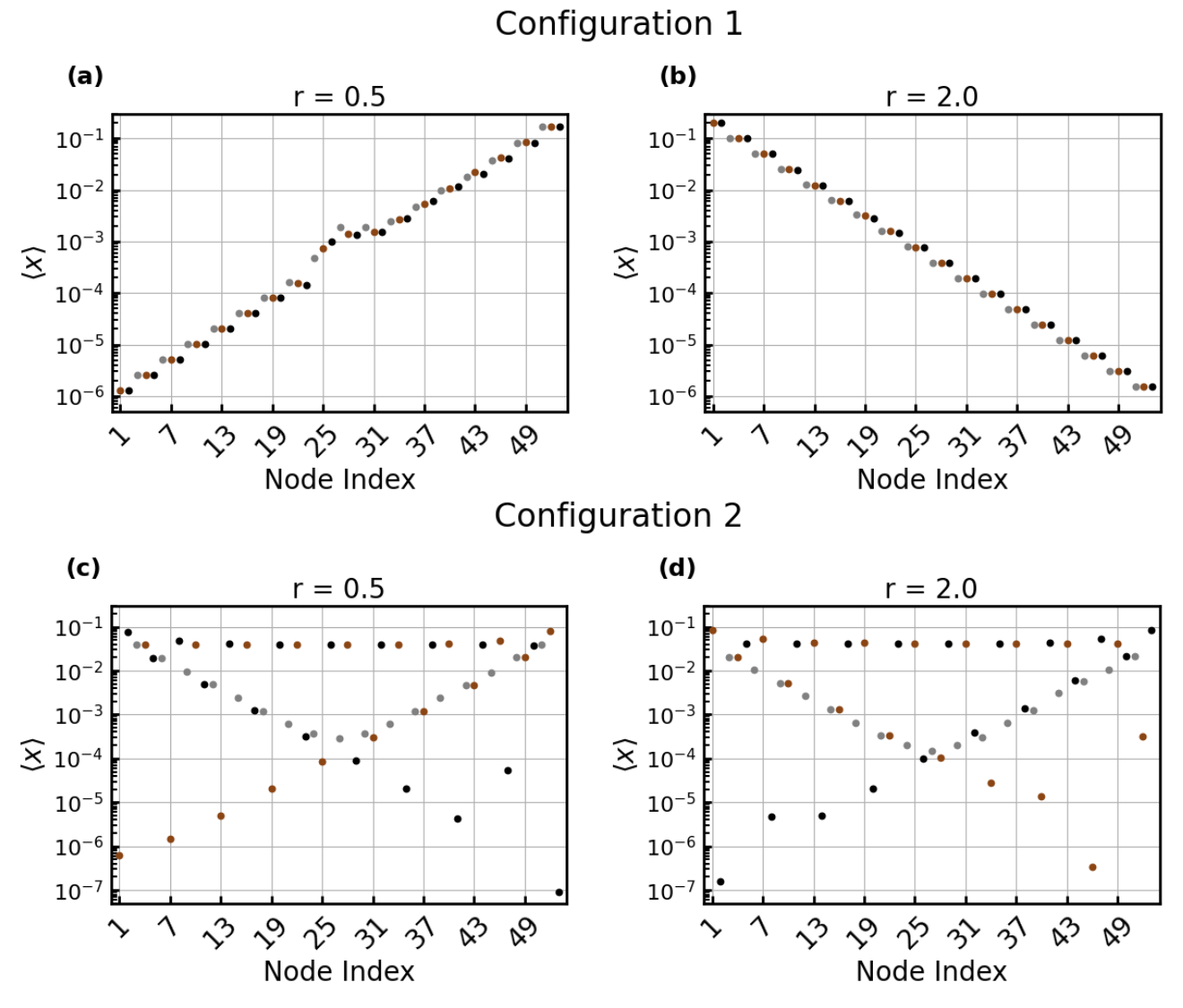}
	\caption{ Results for $N=53$. (a-b) Configuration:1.  Average mass density of each node is reported for $r=0.5$, and $r=2$. The linear shape in semi-log scale confirms the exponential decay from one side to the other side. (c-d) Two opposite corner nodes show the highest depositions depending on $r$. Both alternate red and black nodes exhibit exponential decay. The initial conditions are now taken from the solution $Ax=0$.}
	\label{many_nodes_config_1_25}
\end{figure*}

\begin{figure*}[htbp]
	\centering
	\includegraphics[width= 0.8\textwidth]{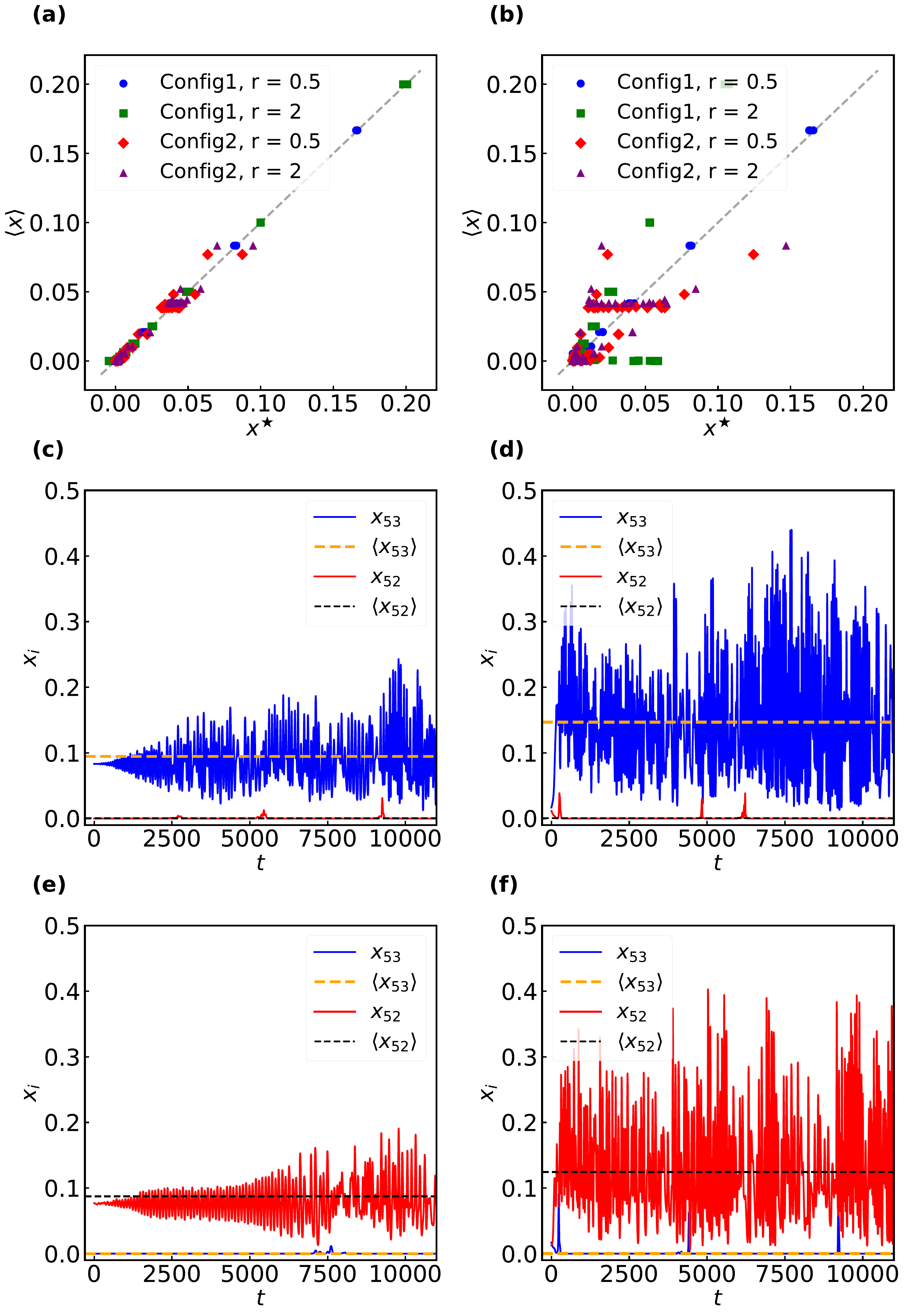}
	\caption{(a-b) $\langle x \rangle _\alpha$   is plotted against  $x^*$. The linear relationship may not hold when random initial conditions are applied (right panel). The network size in this case is fixed at $N=53$. Plotting the time-dependent mass density in (c-d) for $r=0.5$, we have selected two nodes (52 and 53). Dashed lines indicate the long-term average. (e-f)  The opposite situation arises if $r$ is larger than 1. }
	\label{perturbations}
\end{figure*}

First, we elaborate on how $A_1$  can be written using the $A_0^{c1}, A_{-1}^{c1}, A_1^{c1}$. Note that all these matrices are of the order 3X3. So, $\mathbf{0}$ denotes a 3x3 null matrix. 

\bea
A_{1}=
\begin{bmatrix}
A_0^{c1} & A_1^{c1} & \mathbf{0} & \mathbf{0} & \dots & \mathbf{0} & \mathbf{0} & A_{-1}^{c1} \\
A_{-1}^{c1} & A_0^{c1} & A_1^{c1} & \mathbf{0} & \dots & \mathbf{0} & \mathbf{0} & \mathbf{0} \\
\mathbf{0} & A_{-1}^{c1} & A_0^{c1} & A_1^{c1} &  \dots & \mathbf{0} &\mathbf{0} & \mathbf{0} \\
\vdots & \vdots & \vdots   & \vdots & \vdots & \ddots & \vdots & \vdots   \\
\mathbf{0} & \mathbf{0} & \mathbf{0} & \mathbf{0} & \dots & A_{-1}^{c1} & A_0^{c1} & A_1^{c1} \\
\end{bmatrix}
\nonumber
\eea
Thus, $A_1$ is a circulant matrix and in compact form can be written as $A_1 = Circ (A_0^{c1}, A_1^{c1}, \mathbf{0}, \mathbf{0},..\mathbf{0}, A_{-1}^{c1})$.
\newline
The corresponding anti symmetric matrix for configuration 2 is as follows:
\bea
A_{2}=
\begin{bmatrix}
0 & 0 & r_3 & -r_1 & 0 & 0 & 0 & 0 & \dots & 0 \\
0 & 0 & r_2 & 0 & -r_1 & 0 & 0 & 0 & \dots & 0\\
-r_3 & -r_2 & 0 & r_2 & r_3 & 0 &  0 & 0 & \dots & 0\\
r_1 & 0 & - r_2 & 0 & 0 &  r_3 & -r_1 & 0   & \dots & 0 \\
0 & r_1& -r_3 & 0 & 0 & r_2 & 0 & -r_1 &  \dots & 0\\
0 & 0 & 0 & -r_3 & -r_2 & 0 & r_2 & r_3 & \dots & 0 \\
\vdots & \vdots & \vdots   & \vdots & \vdots & \ddots & \vdots & \vdots   & \vdots & \vdots  \\
0 & 0 & 0 & 0 & 0 & 0 & \dots & 0 & -r_2 & -r_3\\
0 & 0 & 0 & 0 & 0 & 0 & \dots & r_2 & 0 & 0\\
0 & 0 & 0 & 0 & 0 & 0 & \dots & r_3 & 0 & 0\\
\end{bmatrix}
\nonumber
\eea
Similarly for configuration 2 we can write, $A_2 = Circ (A_0^{c2}, A_1^{c2}, \mathbf{0}, \mathbf{0},..\mathbf{0}, A_{-1}^{c2})$ and the corresponding sub-matrices are,

\[A_0^{c2} = \begin{pmatrix}
0 & 0 & r_3\\
0 & 0 & r_2\\
-r_3 & -r_2 & 0
\end{pmatrix}, 
A_1^{c2} = \begin{pmatrix}
-r_1 & 0 & 0\\
0 & -r_1 & 0\\
r_2 & r_3 & 0
\end{pmatrix} = (-A_{-1}^{c2})^T.
\]

Therefore the Hamiltonian matrix $\Tilde{H}(k)$ for configuration 2 has the form,
\begin{equation}
\Tilde{H}_{c2}(k) =
\begin{pmatrix}
  2r_1\sin(k) & 0 & ir_3 - ir_2e^{-ik}\\
    0 &  2r_1\sin(k) &  i r_2 - i r_3 e^{-i k}\\
    -ir_3 + ir_2e^{ik} &  -i r_2 + i r_3 e^{i k} & 0\\
\end{pmatrix}
\nonumber
\end{equation}
The behaviour of energy bands and winding number of the above Hamiltonian is discussed in the main text.

\section{Results in large network}
\label{app3}
Using a relatively large graph ($N=53$), we have examined the mass density, confirming our observations in the main text. Fig.\ \ref{many_nodes_config_1_25} displays the outcomes for both configurations when $r=2$ and $r=0.5$ are taken into account. We have selected two sets of initial conditions (IC) in order to further validate the linear relationship ($\langle x \rangle _\alpha \sim x^*$): (I) From the solution $Ax=0$, IC is selected. In two random nodes,  a minor perturbation ($x_i^* \pm 0.001$) is used. (II)  Initial states are purely random. Keep in mind that we have kept the starting point at $\sum_{\alpha = 1}^{S} x_{\alpha}(t=0) = 1$ for both scenarios. We have plotted $\langle x \rangle _\alpha$  with respect to  $x^*$ in the Fig.\ \ref{perturbations}(a) (case I). In Figure \ref{perturbations}(c), the time signals of two nodes (53 and 52) are displayed (configuration II). Node 53 exhibits irregular oscillation (blue), while Node 52 (red line) remains near zero. In this case, $r$ is set to 0.5. The average is shown by the dashed horizontal lines (after discarding the transient). When $r$ is set to 2, the opposite situation transpires (node 52 exhibits irregular oscillation and 53 remains near zero). The results for case II, in which the IC are chosen at random, are shown in Fig.\ \ref{perturbations}(b), (d), and (f). Random ICs have the effect of increasing the amplitudes of the time signals (node 53 in (d), node 52 in (f)). As a result, there is some disruption in the linear relationship between $\langle x \rangle _\alpha$ and $x^*$.
Hence, a careful selection of initial conditions is necessary to observe the exponential decay (or strong deposition of mass density in the opposite corner). The phase transition happens at $r=1$. Starting from random ICs, it has been numerically observed that (not shown here), for values of $r$ close to 1, the mass density of the configuration will be in better agreement with the $Ax=0$ solution.

\end{document}